\documentclass[epj]{svjour}
\usepackage{graphics}
\input epsf

\begin{document}
\title{A microscopic mechanism  
for rejuvenation and memory effects in spin glasses}

\author{S. Miyashita\inst{1} 
\and E. Vincent\inst{2}
}                     
\offprints{miya@yuragi.t.u-tokyo.ac.jp, \\ vincent@drecam.saclay.cea.fr }          
\institute{Department of Applied Physics, University of Tokyo, 7-3-1
Bunkyo-ku, Tokyo 113-8656, Japan  
\and 
Service de Physique de l'Etat Condens\'e, CEA Saclay, 91191 Gif sur
Yvette Cedex, France}
\date{Received: date / Revised version: April 27, 2001}
%
\abstract{
Aging in spin glasses (and in some other systems) reveals astonishing
effects of `rejuvenation and memory' upon temperature changes. In this
paper, we propose microscopic mechanisms (at the scale of spin-spin
interactions) which can be at the origin of such phenomena. Firstly,
we recall that, in a frustrated system, the {\it effective average
interaction} between two spins may take different values (possibly
with opposite signs) at different temperatures. We give simple
examples of such situations, which we compute exactly. Such mechanisms
can explain why new ordering processes ({\it rejuvenation}) seem to
take place in spin glasses when the temperature is lowered.  Secondly,
we emphasize the fact that {\it inhomogeneous interactions} do
naturally lead to a wide distribution of relaxation times for
thermally activated flips. `Memory spots' spontaneously appear, in the
sense that the flipping time of some spin clusters becomes extremely
long when the temperature is decreased. Such memory spots are capable
of keeping the {\it memory} of previous ordering at a higher
temperature while new ordering processes occur at a lower
temperature. After a qualitative discussion of these mechanisms, we
show in the numerical simulation of a simplified example that this may
indeed work. Our conclusion is that certain chaos-like phenomena may
show up spontaneously in any {\it frustrated} and {\it inhomogeneous}
magnetic system, without impeding the occurrence of memory effects.
}

\PACS{
{75.50.Lk} {Spin glasses and other random magnets} \and
{75.10.Nr} {Spin-glass and other random models}
}

\maketitle

\section{Introduction}

The phenomena of slow dynamics in spin glasses, well known from the
experiments in which the out-of-equilib\-rium effects are prominent
\cite{Sitges,Nordblad}, have been these last years the subject of
significant developments in theory \cite{Review}, numerical
simulations \cite{numerics}, and also experiments on other glassy
systems like e.g. polymers \cite{PMMA}, supercooled liquids
\cite{Nagel}, dielectrics \cite{Levelut} or gels \cite{gel}. In the
spin-glass phase, dynamical response functions evolve with time: this
is the {\it aging} phenomenon, comparable with {\it physical aging}
which has been widely studied in the rheology of glassy polymers
\cite{Struik}. Aging in spin glasses is evidenced in two general
classes of {\it ac} and {\it dc} experiments. The starting point of
aging is when the spin glass is cooled from above $T_g$ down to some
temperature $T_1$ (usually $\sim 0.5-0.9T_g$).

In $dc$ experiments, e.g. zero-field cooled (`ZFC') procedure, the
sample is cooled in zero field and kept at $T_1$ during a waiting time
$t_w$. After $t_w$, a weak magnetic field is applied, and the slow
increase of the magnetization is recorded as a function of time
$t$. The response curves obtained depend on both independent time
variables $t$ and $t_w$: they become slower and slower for increasing
$t_w$.  Similar results are obtained in the {\it inverse} procedure of
cooling in a weak field and removing the field after $t_w$ (`TRM'
procedure). In $ac$ experiments, equivalently, the response to a small
oscillating field at frequency $\omega$ is found to relax slowly as
the time $t$ from the quench elapses: here again two separate time
scales ($\omega^{-1},t$) are involved \cite{Sitges}. Such aging effects
have now been clearly identified in numerical simulations of the
three-dimensional Edward-Anderson model, and are also found in the
analytical treatment of some mean-field models \cite{Review,CuKu}. They
appear as a characteristic feature for the dynamics of randomly
interacting objects.

The effect on aging of temperature changes has led to some non-trivial
experimental results \cite{hierarki,DTupps}, recently emphasized as
`memory and chaos' or `memory and rejuvenation' effects
\cite{memchaos}. Figure 1 (from Ref.\cite{VD}) shows the
characteristic example of a negative temperature cycling experiment in
{\it ac} mode. After a long aging stage at $T_1=0.72T_g$
(characterized by a downwards relaxation of $\chi"(\omega)$ by about
30\%), a further cooling to $T_2=0.54T_g$ results in an apparent {\it
reinitialization} of aging (`rejuvenation'): $\chi"$ rises up to about
the value that would be obtained after a direct quench, and a strong
relaxation restarts.  This observation contradicts the expectation
from simple thermal slowing down, and suggests a possible indication
of `chaos in temperature' as proposed in
Refs.\cite{bray,domains}. Following this `rejuvenation' effect upon
cooling, a completely different phenomenon is observed when re-heating
from $T_2$ to $T_1$. The {\it memory} of aging at $T_1$ is retrieved,
in the sense that $\chi"$ goes back to the value attained at $T_1$
before the temperature cycle, and relaxes in continuity with the
previous part \cite{hierarki}.

\begin{figure} \label{fig1}
\center\epsfysize=7cm\epsfbox{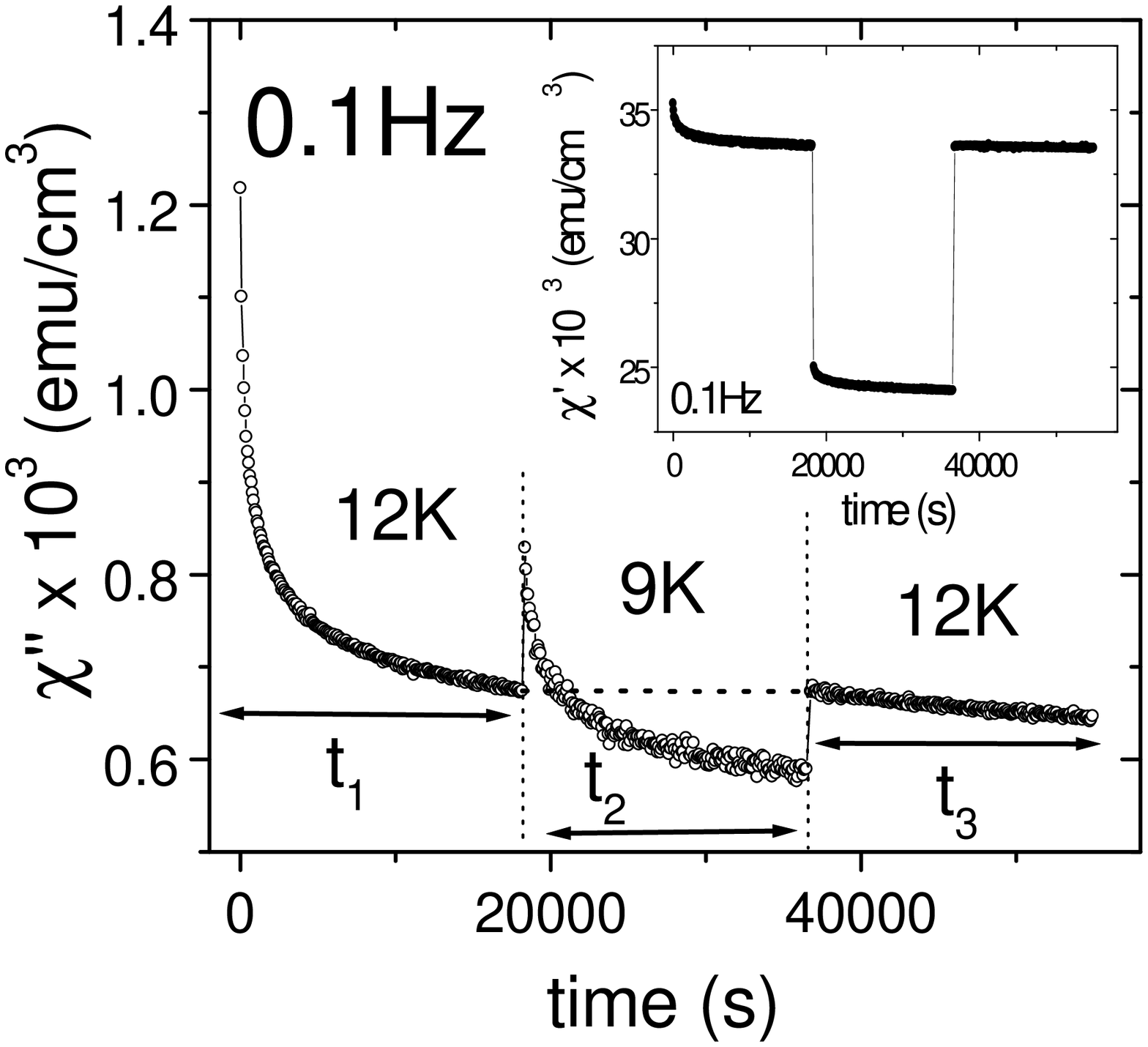} 
\caption{Effect on the $\chi''$ relaxation of a negative temperature
cycling ($CdCr_{1.7}In_{0.3}S_4$ insulating spin glass with
$T_g=16.7$K, frequency 0.1Hz, from Ref.\cite{VD}). Aging is
mostly reinitialized during negative cycling (rejuvenation). The inset
shows the $\chi'$ behaviour during this procedure: the same effects
are visible, but less clearly.}
\end{figure} 

In the same way, a measurement of $\chi"(T)$ during continuous
re-heating from low temperatures shows a `dip' centered around $T_1$
\cite{memchaos}. Astonishingly, it is even possible to `imprint' and
`read' several memories at different temperatures, like in the example
presented in Fig.2 \cite{VD}. However, we shall mainly address here
the question of the basic mechanisms underlying the observations of
Fig.1. The multiple memories of Fig.2 will only be discussed as a
possible extrapolation of our approach in Section 4.

\begin{figure} \label{fig2}
\center\epsfysize=6cm\epsfbox{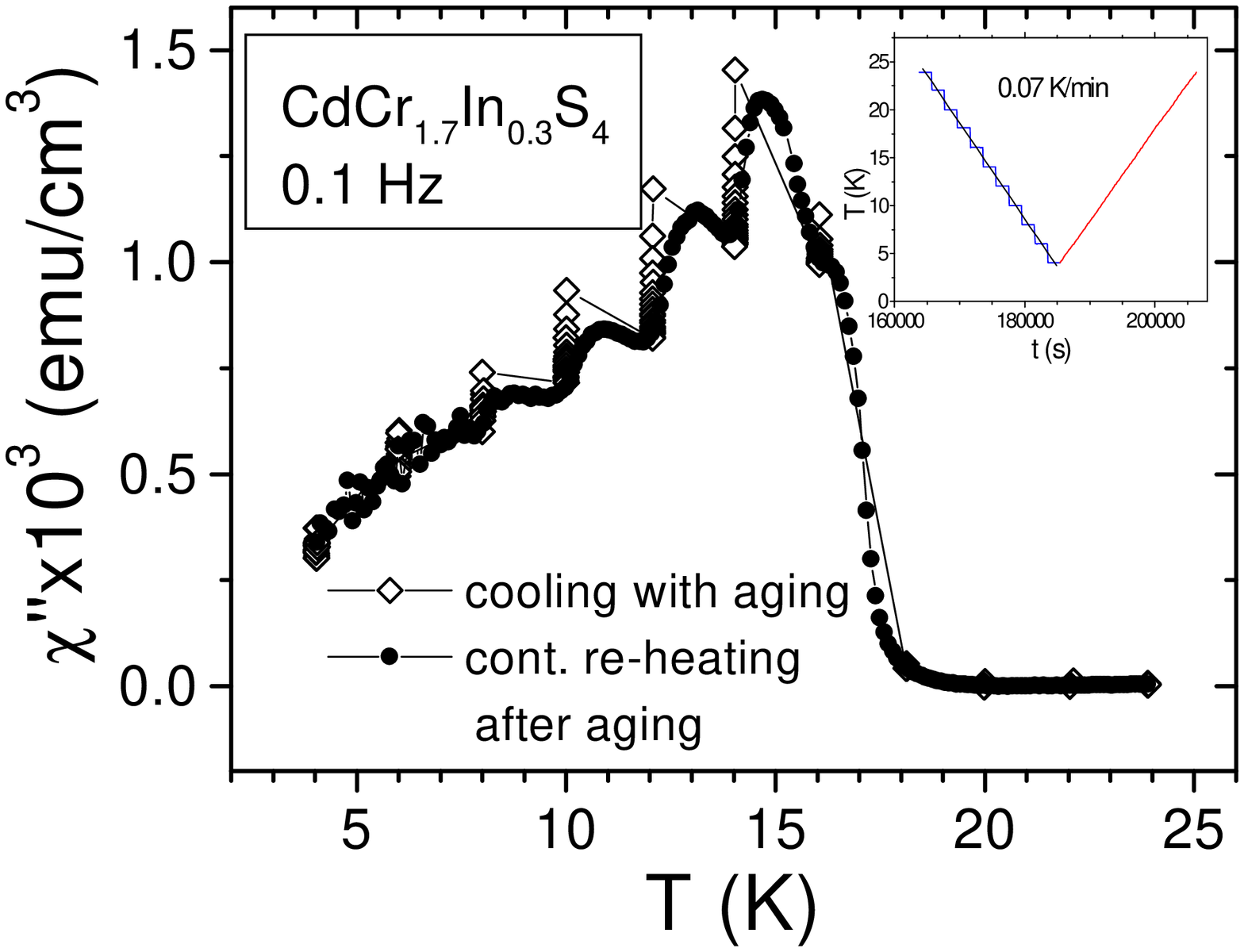}
\caption{An example of multiple `rejuvenation and memory' steps. The
sample was cooled by 2K steps, with an aging time of 2000 sec at each
step (open diamonds). Continuous re-heating at 0.001K/s (full circles)
shows memory dips at each temperature of aging (from Ref.\cite{VD}).}
\end{figure} 

A comprehensive description of this class of experiments in terms of a
{\it hierarchical organization} of the {\it meta\-stable states} as a
function of temperature has been developed, and gives a satisfactory
account of all results \cite{hierarki}. This phenomenological picture
has been made quantitative in models of random walk in a hierarchical
set of traps \cite{BD,Nemoto}.  However, the interpretation of the
rejuvenation and memory effects in the {\it real space} of spins
remains puzzling. From a comparison with aging in disordered
ferromagnets, a scenario of {\it pinned wall reconformations}
\cite{DWhierarchy} has been proposed, in which the hierarchy of
reconformation length scales is a real space transcription of the
hierarchy of states \cite{ferro}. But the exact nature of such walls
in a spin glass remains mysterious \cite{martin}. It is the purpose of
the present paper to take a different point of view, and to consider
{\it microscopic} and {\it explicit} mechanisms, at the scale of
spin-spin interactions, which are possible candidates for being at the
origin of these memory and rejuvenation effects.

On the basis of previous works on the reentrant phenomena in
frustrated systems\cite{REENT}, we study example situations in which a
slow evolution towards equilibrium at a certain temperature can be
irrelevant to equilibration at another temperature ({\it rejuvenation}).  On
the other hand, the {\it memory} effect implies that spin rearrangements at
one temperature do not irreversibly affect the structure grown at
another temperature. 
This is the case for the mechanisms which will be considered here. 
Due to the inhomogeneity of the interactions, intricate
developments of the spin-spin correlation take place, which should
play an important role in the peculiar properties of spin glasses.
Let us note that this intricate structure of the domains growing at
different temperatures does not necessarily imply a fractal geometry.

In this paper, we limit ourselves to the study of examples of
frustrated magnetic structures which, on the basis of phenomenological
arguments, can be shown to reproduce the `single memory' situation.
The question of the extension of this basic mechanism to a double or
multiple memory case, although conceivable (as discussed in Section
4), remains beyond the scope of the present paper.

\section{Temperature dependent effective interactions and
memory spots}

In spin glasses, the ferromagnetic and antiferromagnetic bonds are
randomly distributed.  For statistical reasons, some regions of the
lattice are highly frustrated, while others are less frustrated.  The
distribution of such regions is at the origin of complicated ordering
processes, which are not as intuitively understandable as in the case
of ferromagnets.

It has been early recognized \cite{Seiji} that, in very simple
frustrated systems of a few spins which can be analyzed exactly, the
effective coupling constant between spins may behave strangely, such
as changing of sign with temperature.  This property, due to
frustrated spin coupling, has been shown to generate reentrant
phenomena\cite{REENT}.  A similar idea has been more recently
developed in the case of the Edwards-Anderson model \cite{Huse}.

As an example, let us consider the simple 3-spin system pictured
in Fig.3(a). If we consider the case where $J_1<0$ (antiferromagnetic) and
$J_2>0$ (ferromagnetic), 
the spins $\sigma_1$ and $\sigma_2$ interact 
by frustrated bond structures.
An effective coupling $J_{\rm eff}(T)$ between 
$\sigma_1$ and $\sigma_2$ can be defined by 
\begin{equation}
e^{\beta J_{\rm eff}(T)\sigma_1\sigma_2} \propto
\sum_{\sigma_0=\pm 1}e^{\beta J_1 \sigma_1\sigma_2+
 \beta J_2 (\sigma_1+\sigma_2)\sigma_0}
\end{equation}
where $\beta = 1/k_{\rm B}T$, and $J_{\rm eff}(T)$ is explicitly given by
\begin{equation}
J_{\rm eff}(T)=J_1+{k_{\rm B}T\over 2}\ln[\cosh(2{J_2 \over k_{\rm B}T})].
\end{equation}
 In the case 
$|J_1| < J_2$, $J_{\rm eff}$ changes sign as a function of 
temperature, as displayed in Fig. 3(b).

\begin{figure} \label{fig3}
\center\epsfysize=8cm\epsfbox{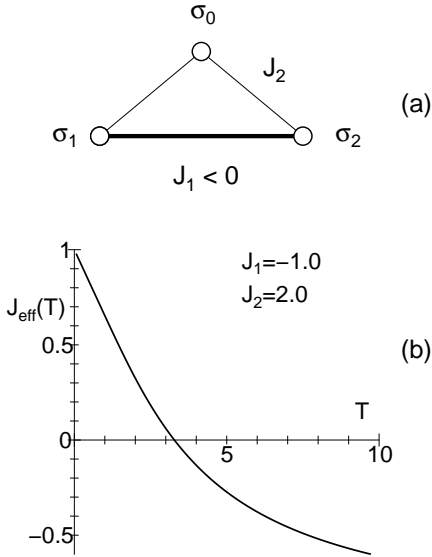}
\caption{(a) An example of frustrated interaction. (b) The effective
coupling constant $J_{\rm eff}(T)$ as a function of  temperature.}
\end{figure} 

From this simple example, it is clear that, due to frustration,
ordering processes can change qualitatively with the temperature. We
have listed in Appendix some realizations of the effective coupling
 in various frustrated situations. In the quoted
examples, we see that the effective interactions may change with
temperature in very different ways, even non-monotonically in some
cases.  If such bond configurations are randomly distributed in the
lattice, it is clear that the equilibrium correlations at a given
temperature do not coincide with those at other temperatures, a
natural mechanism for {\it rejuvenation} effects.  In a recent study
of domain growth processes in a Mattis model \cite{YOSHINO}, the
consequences of {\it bond changes} on rejuvenation effects have been
investigated (see a more detailed discussion in our last section). The
microscopic mechanisms studied in our present paper can be considered
as a possible explanation for the bond changes that have explicitly
been assumed in \cite{YOSHINO}.

Let us note that the effective couplings which are considered here
correspond to a coarse-grained picture of the original lattice. Thus,
the {\it spins} which are interacting via the effective bonds do in
fact represent {\it block spins}, i.e. clusters of spins with less
frustrated bond structures, as proposed in \cite{REENT}. They are
entities which already possess a significant entropy.

We expect the {\it memory} effect to be related to another
characteristic of spin-glasses, which is that the relaxation times of
spins distribute widely from spin to spin due to inhomogeneous
interactions\cite{TAKANO}.  As an example, let us consider the case
depicted in Fig. 4, in which all couplings are ferromagnetic, one of
them ($|J_1|$) being much larger than those of the surrounding bonds
$J_0$ $(J_1\gg J_0)$. 

\begin{figure} \label{fig4}
\center\epsfysize=9cm\epsfbox{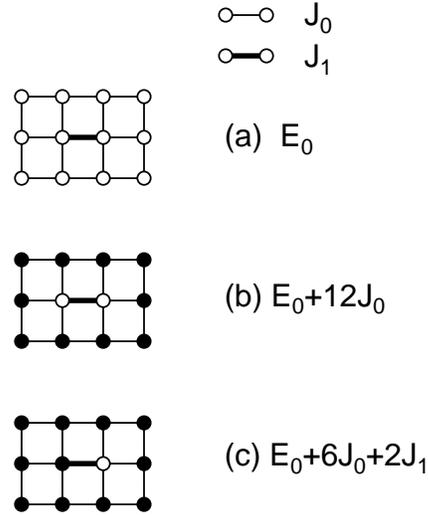}
\caption{ An example of memory spot ($J_1\gg 3J_0$). 
 (a) stable state, (b) metastable state, and (c)unstable state.
In order to relax
from the metastable state (b) to the ground state (a), the system has
to cross an energy
barrier $\Delta
E=2J_1-6J_0$ due to the intermediate state (c).
}
\end{figure} 

In Fig. 4(a), the ground state configuration
with energy $E_0$ is shown; if all the external spins change sign, the
two strongly coupled spins become metastable with energy $E_0 +
12J_0$, Fig. 4(b).  In order to relax the two spins to the ground
state, the system has to pass an intermediate state of higher energy
$E_0 + 2J_1 + 6J_0$, Fig. 2(c).  This corresponds to an energy barrier
$\Delta E=2J_1-6J_0$.  Thus, if $T\ll \Delta E$, it is difficult to
flip the two spins, even if the surrounding spins are changed.  The
relaxation time is
\begin{equation}
\tau_{\rm mem}= \tau_0 \exp\left({\Delta E\over T}\right),
\end{equation}
where $\tau_0$ is a microscopic time; $\tau_{\rm mem}$ becomes
suddenly long below a certain temperature.  We expect that this kind
of strongly coupled clusters are effectively realized in the less
frustrated regions of a spin glass.  Such clusters are distributed in
the system, and can {\it memorize} a grown pattern of ordered
configuration at a given temperature. We call them `memory spot'.  
When the temperature is lowered, the magnetizations of the stronger
memory spots, 
which memorize the configurations at higher temperatures, are stable.  
In turn, when the temperature is raised, the memories
for the lower temperatures are erased.

\section{Rejuvenation and memory effects from basic mechanisms}

Let us now summarize how a `rejuvenation and
memory' scenario can be built up with temperature dependent {\it effective
interactions} and {\it memory spots}. Our 
qualitative discussion will be followed by the Monte-Carlo simulation
of an example lattice.

If we quench the system from a high temperature to a certain value,
say $T_1$, order corresponding to the minimization of the effective
interaction energies at this temperature begins to grow.
Let us consider the example sketched in Fig. 5(a).  
The fuzzy lines correspond to regions of high frustration, where the 
effective bonds at $T_1$ are weaker than in other parts and where in
consequence domain walls are easily trapped.
The `configurational domains' delimited by these lines represent 
less frustrated regions.  

Here we suppose that each domain has two degenerate minimum energy
states, which are denoted by up and down arrows. 
We assume that the minimum energy state of the whole system 
with respect to the effective interactions at $T_1$ corresponds 
to all domain arrows pointing in the same direction.
As we discuss later, the memory spots follow the
direction of the domains in the time evolution at $T_1$, 
and record their direction at lower temperatures. 
In Fig. 5(a), the magnetization direction of the memory spots is  
shown by arrows for the equilibrium state.

\begin{figure} \label{fig6}
\center\epsfysize=4.0cm\epsfbox{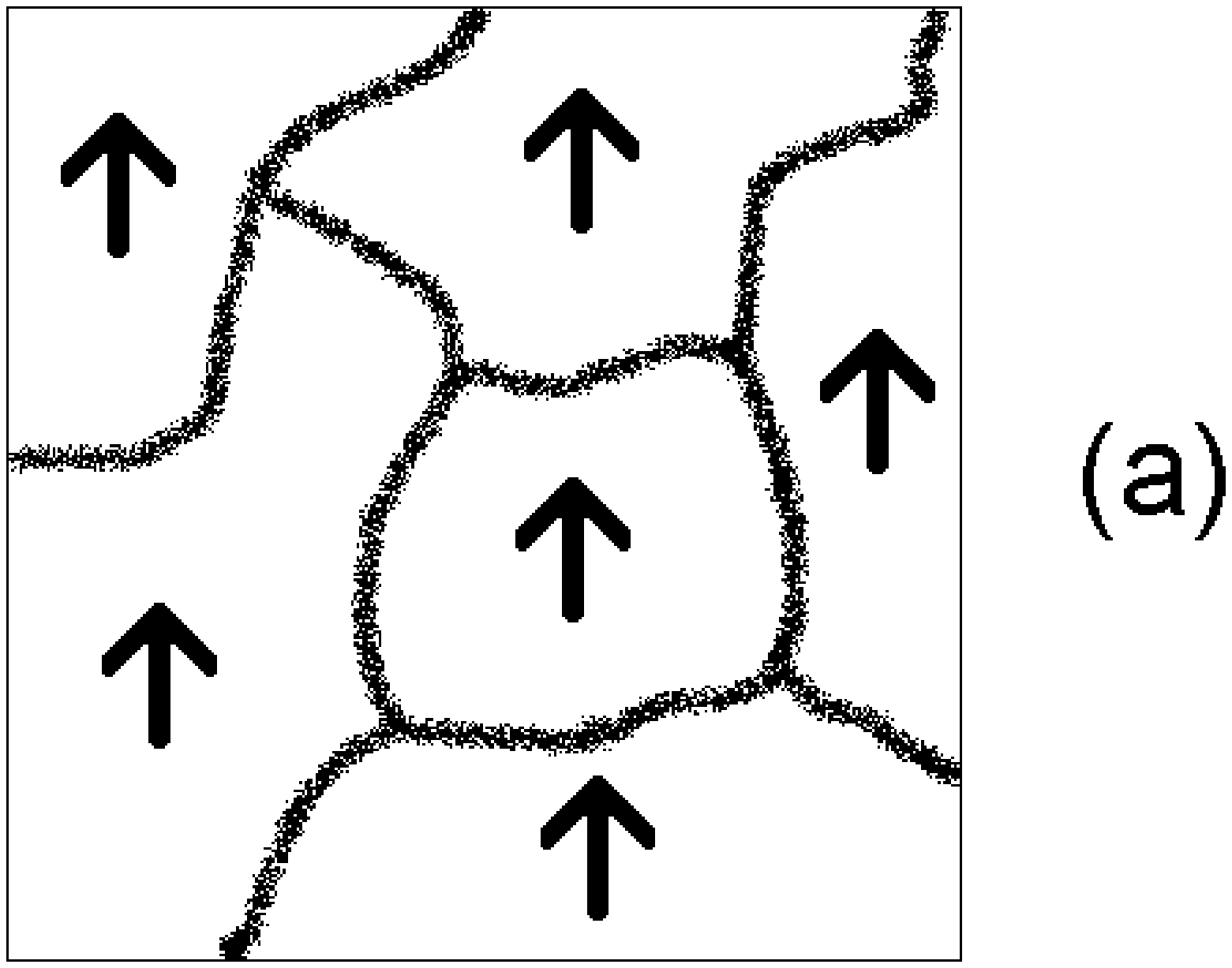}
\center\epsfysize=4.0cm\epsfbox{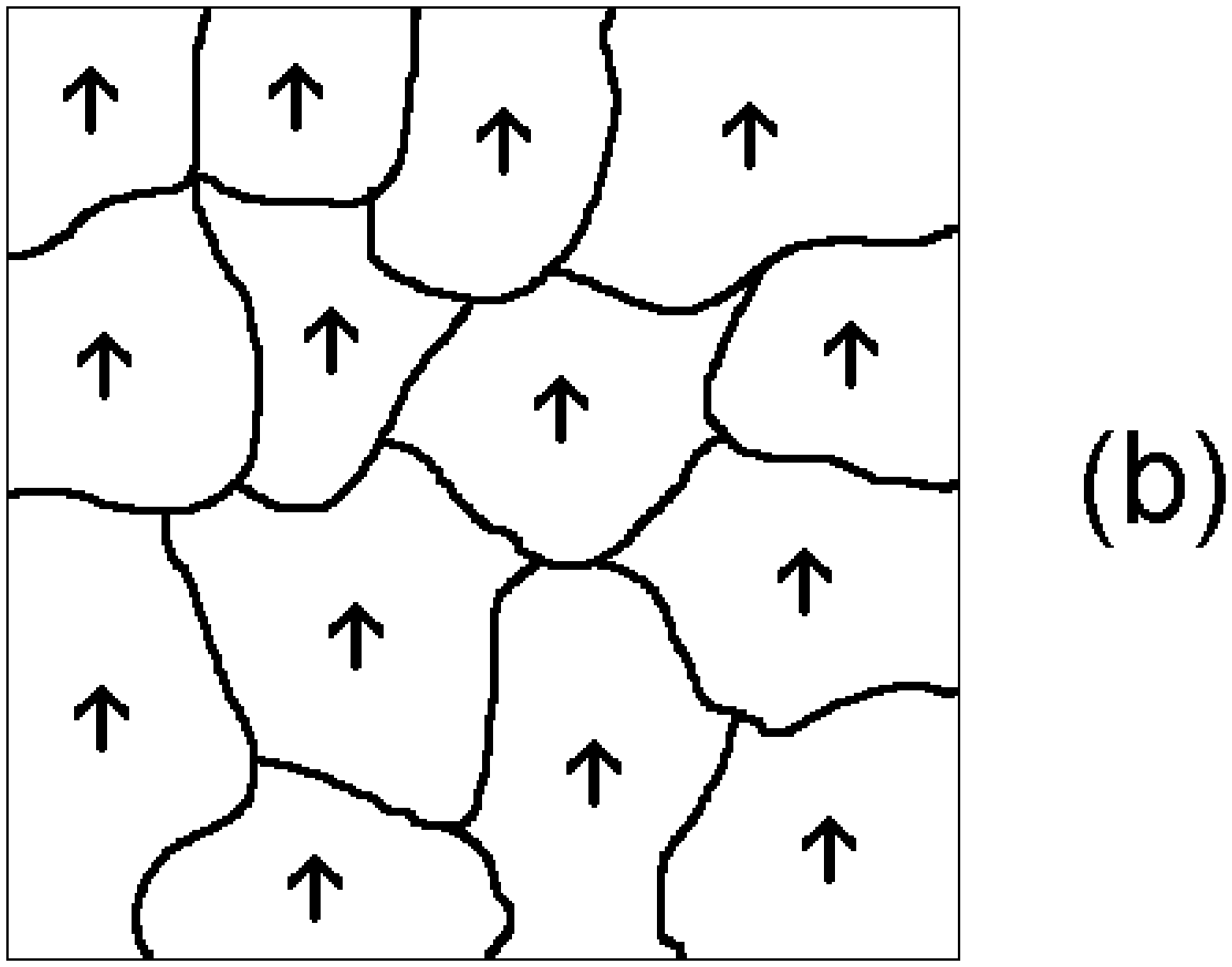} 
\caption{Schematic domain configurations after a waiting time  (a) at $T_1$ and 
(b) at $T_2$. The domain boundaries, which consist of
weaker bonds where the domain wall is easily trapped, are represented
by lines. The memory spots (in (a) and (b)) are shown by arrows.
The equilibrium spin correlations inside a domain can be rather general
(Mattis model like); they do not coincide at both temperatures (the
{\it up} direction of the arrows refers to the projection onto an
arbitrary two-fold state).
 }
\end{figure}

In a short time $\tau_{\rm iD}$ after quenching, local order has been
realized within each domain (low frustration). But no order between
the domains has yet been established.  That is, at this stage, the
arrows of the domains are independently oriented.  Then, waiting
during $t_w$, correlations {\it among} the domains develop, with a
flipping time scale $\tau_{\rm DD}$; meanwhile, $\chi"(\omega)$
decreases. The increase of the correlation length is very slow because
as mentioned above the domain walls are naturally pinned at the
boundaries \cite{slowDG}.

We assume that the flipping time $\tau_{\rm mem}$ of the memory spot
is less than the flipping time $\tau_{\rm DD}$ of the $T_1$
configurational domains.  Hence the memory spots follow the direction
of the domains, and record their direction as order among the domains
develops.  At lower temperatures, the magnetization of these memory
spots will be frozen.

Then we change the temperature to $T_2<T_1$.  Fig. 5(b) is a sketch of
the configurational domains at $T_2$.  Since the values of the {\it
effective} bonds have changed, the grown domain structure grown at
$T_1$ is just a random configuration for $T_2$ (rejuvenation). Thus
new domains relevant to $T_2$ start growing, and $\chi"(\omega)$ rises
back to a higher value.

Although the most part of the lattice has been rejuvenated, the memory
spots remain stable because they consist of unfrustrated structures
and $\tau_{\rm mem}$ increases rapidly as the temperature goes down.
The structure at $T_2$ will be also recorded in smaller memory spots.
In this way, the structure of order at each temperature can be
recorded by memory spots of proper size,
whose magnetization remains frozen at lower temperatures.

When re-heating to $T_1$, the order developed at $T_2$ in most of the
lattice is now random with respect to order at $T_1$. However, in each
$T_1$ configurational domain, the memory spot has memorized the
previous direction of order.  Therefore the domains tend to re-order
according to their memorized direction, which rapidly reconstructs the
configuration obtained at $T_1$ before the temperature cycle.  Thus,
in a very short time, $\chi"(\omega)$ decreases back to the value
previously obtained (memory).

Let us now illustrate the above scenario by the numerical simulation
of a simple model. We consider a lattice made of 5x5 times the unit
clusters ($10\times 10$) shown in Fig. 6.  The bonds $J_1$ and $J_2$
will mark the boundaries of the configurational domains at
respectively $T_1$ and $T_2$.  The memory spots are given by $J_3$.
All other bonds are $J_0$. Table 1 displays the bond values at
$T_1=2.0$ and $T_2=1.0$, the temperature variation being assumed to be
due to the mechanisms described above.

\begin{figure} \label{fig7}
\center\epsfysize=8cm\epsfbox{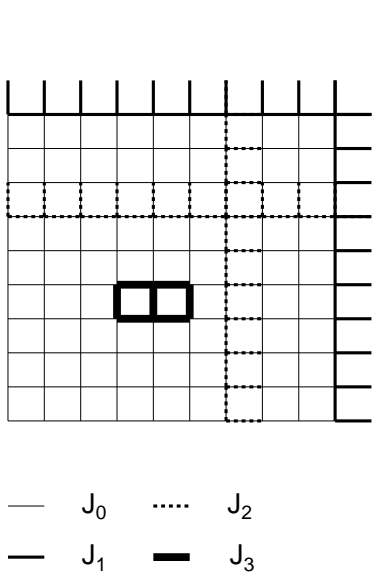}
\caption{Bond configuration (unit). The boundaries of the 
`configurational clusters' will be given by the
bold solid lines at $T=T_1$, and by the dotted lines at $T=T_2$}
\end{figure}

\begin{table}
\begin{center}
\begin{math} 
\begin{tabular} {|c|cccc|}
\hline
 $T$ & $J_0$ & $J_1$ & $J_2$ & $J_3$ \\ \hline
 $T_1$= 2 & 1 & 1 & 0.5 & 3.2 \\ \hline
 $T_2$=1 & $-1$ & $-0.5$ & $-1$ & $3.2$ \\ \hline
\end{tabular}
\end{math}
\caption{Coupling constants for the bonds in Fig.6.}
\end{center} 
\end{table}

Obviously, from Table 1, the order is ferromagnetic at $T_1$ and
(mostly) antiferromagnetic at $T_2$. We have chosen these two
equilibrium configurations at $T_1$ and $T_2$ as simple examples; they
might as well be any ground state choice of a Mattis model, as has
been used in \cite{YOSHINO}, as far as they are sufficiently different
from each other.

\begin{figure} \label{fig8}
\center\epsfysize=12cm\epsfbox{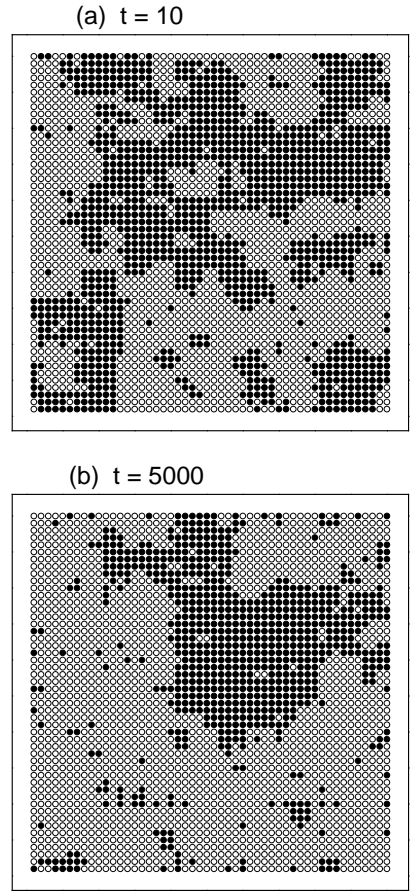}
\caption{
(a) a configuration just after the short range order developed in each domain at $T_1$, 
(b) domain structures after a certain time at $T_1$, where
correlation among the domains developed.
}
\end{figure}

\begin{figure} \label{fig9}
\center\epsfysize=16cm\epsfbox{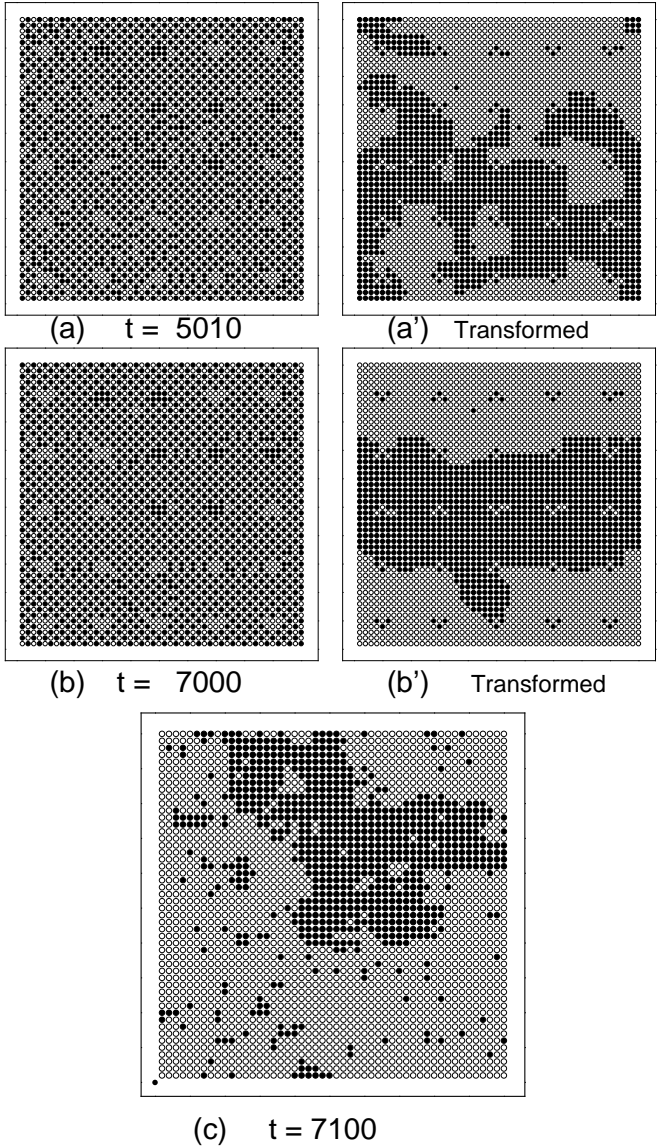}
\caption{
(a) an early stage configuration after the change of the temperature
to $T_2<T_1$,
(a') configuration displayed in the antiferromagnetic gauge 
$\sigma_{i}\rightarrow \sigma_i \times \sigma^0_i$, where $\sigma^0_i$
gives the antiferromagnetic order, 
(b) developed new domain structure, at $T_2$, 
(b') configuration displayed in the antiferromagnetic gauge 
and
(c) recovery of the previous domain structure after the temperature comes back to $T_1$.}
\end{figure}

The initial configuration is completely random.  In a short time
$\tau_{\rm iD}$, order inside the domain develops, and the correlation
length increases up to the size of the unit cluster ($\sim$ 10), as
shown in Fig.7(a) for $t=10$MCS. In a second stage, order among the
domains develops at the time scale $\tau_{\rm DD}\gg \tau_{\rm iD}$
(Fig.7(b) at $t=5000$MCS).

Then, at $t=5001$MCS, we change the temperature to $T_2$, which means
changing the values of the effective couplings (Table
1). Antiferromagnetic order is now favoured.  At $t=5010$MCS
(Fig.8(a)), the new structure appears random (rejuvenation), but the
memory spots are visible, clearly keeping track of previous
ordering. In order to emphasize the new ordering process at $T_2$, we
projected the configuration attained at $t=5010$MCS (Fig.8(a)) onto an
{\it antiferromagnetic} ground state (`transformed' picture), as
represented in Fig.8(a').  The new order among the new clusters has
not yet grown.  The $T_1$ memory spots can also be seen.

A new domain structure starts to develop within the time scale of
$\tau_{\rm DD}(T_2)$.  Fig. 8(b) (direct) and 8(b') (transformed) show
the development at $t=7000$MCS of the new order at $T_2$ (the memory
spots are still visible).

Then, at $t=7001$MCS, the temperature is increased back to its
original value $T_1$. In a short time, most of the previous ordering
is recovered, as can be seen from Fig.8(c) at $t=7100$MCS. The
comparison of Fig.8(c) with Fig.7(b) shows that the memory of ordering
at $T_1$ has actually survived to the rejuvenation process at
$T_2$. The rapid re-ordering processes which occur immediately after
re-heating to $T_1$ are likely to produce a short $\chi''$ `transient
relaxation', which has indeed been noted in some experiments
\cite{hierarki,DTupps}.

\section{Discussion}

\subsection{A plausible scenario for multiple memories ?}
Beyond this simple example of rejuvenation and memory phenomena at
{\it two} temperatures, the possibility of imprinting and reading {\it
multiple memories} at several temperatures using the above mechanisms
is a very incentive issue, even if it remains somewhat speculative.
At different temperatures which are separated by a large enough
interval $\Delta T$, we consider that the ordering patterns are
decorrelated from each other, like different ground states in a Mattis
model for different sets of bonds (for smaller $\Delta T$'s, it is
clear that the ordering patterns will have similarities, and that the
memory spots cannot fully work).

We suppose that the bond configuration of the memory spots is frozen
at lower temperatures, as well as their orientation. Hence they
memorize the signs of the local order (which is two-fold at each
temperature).  They will play the role of nucleation centers when the
temperature is raised back.

The important point is that the memorized patterns do not correspond
to ordering at other temperatures.  At lower temperatures, the memory
spots from higher temperatures are like {\it frozen impurities}.  That
is, the memory spots of a given pattern cannot act as nucleation
centers for another pattern.  This happens effectively in our
numerical example (Fig.8): the ferromagnetically ordered spots do not
have a significant influence on the antiferromagnetic order.

In the previous section, we chose ferromagnetic (F) and
antiferromagnetic (AF) order as examples of two independent ordering
patterns at $T_1$ and $T_2$.  If we now consider a third temperature
$T_3$ ($T_3<T_2<T_1$), the ordering pattern at $T_3$ must be
independent of both F and AF, as mentioned above, and the memory spots
for $T_1$ (F-ordered clusters) and for $T_2$ (AF-ordered clusters)
will just be like frozen impurities.  One may think of repeating this
process in further and further cooling.  How many ordering patterns
can be treated as independent ones is an interesting problem,
which is related to the question of pattern recognition in random
networks, as studied in the Hopfield model \cite{network}.
In real systems, the number of spins is quite large, and 
it is not difficult to have several {\it independent} patterns; 
the present mechanism may thus work to memorize successive {\it independent}
patterns at different temperatures. 
But the memory spots must be large enough to distinguish between different
patterns, and also they have to be small compared with the ordering
domains. In the simulation of the previous section, the ordering
domains were $10\times 10$, and the memory spots were $2\times 3$. It
is clear that the demonstration of multiple memories by simulations of
our scenario will strongly be hindered by size limitations.

\subsection{Rejuvenation effects and chaotic behaviour}

The mechanism proposed here to be at the basis of rejuvenation effects
is of `chaos-type' \cite{bray,domains}, in the sense that it provides
a microscopic basis for a `chaotic' temperature dependence of the
bonds.  Memory is obtained due to the `memory spots' which appear
spontaneously in any inhomogeneous system. The question of `memory
despite rejuvenation' in {\it domain growth processes} has recently
been discussed by Yoshino et al \cite{YOSHINO} in an analytical and
numerical study of the Mattis model.  The Mattis model is purely
random, but with no frustration, and the doubly degenerate ground
state can be {\it arbitrarily chosen} by the set of the magnetic
interactions.  In \cite{YOSHINO}, the bonds are changed `by hand' from
one set to another, and back (this is not far from the numerical
example that we have presented here). A first order grows, say of
A-type, then some other B-order develops, and coming back to A-bonds
one may examine how far A-order has been preserved despite the
rejuvenation caused by the growth of B-order. In the Mattis model,
domain growth is a fast process because there is no frustration, so
the memory of A-domain growth is rapidly erased by the growth of
B-type domains. Due to its simplicity, the Mattis model is a useful
`toy-model' which even allows some analytic calculations
\cite{YOSHINO}.  In a real (fully disordered) system, the growth of
the correlation length is naturally much slower.
 
The study of the Mattis model \cite{YOSHINO} shows that the large
scale shape of the A-domains is preserved (memory), while the effect
of rejuvenation can be seen as small scale `holes' in the big
A-domains.  The same features can be observed in our numerical
example, comparing Fig.7(b) and 8(c), but memory is here more robust
thanks to the memory spots.

The fact that memory is preserved in {\it large} length scales, while
rejuvenation occurs at {\it short} length scales, agrees well with the
hierarchical sketch corresponding to the experiments
\cite{hierarki}. In contrast, the usual discussion of chaos in spin
glasses \cite{bray,domains} is in terms of an overlap length {\it
beyond which} the equilibrium correlations are re-shuffled by a
temperature or bond change. In \cite{YOSHINO}, and to a certain extent
(apart from the memory spots) in the present numerical example, the
overlap length between both considered states is zero, and
rejuvenation and memory occur at respectively short and large length
scales in a `hierarchical' fashion.

The question of `chaos' \cite{bray} in the Edwards Anderson spin glass
is rather puzzling. The spins in our present simplified picture can
perhaps be compared with `block spins' in the Edwards Anderson model,
which have been shown in \cite{Huse} to interact via effective bonds
of temperature-\-dependent signs (in the same spirit as in
\cite{REENT,Seiji}.  It is intriguing that, in the present numerical
simulations of the Edwards Anderson spin glass \cite{numerics}, no
clear rejuvenation and memory effects could be found in the dynamics
(as well as no sign of chaos in the statics). In our present scenario,
temperature dependent interactions and memory spots are obtained from
simple bond arrangements.  In a real spin glass, we argue that similar
situations should statistically be realized due to the high number of
random bonds. It is likely that this is not the case for numerical
simulations, in which the number of spins may remain too small to
allow the presence of such complicated structures as proposed here. On
the other hand, the time scale of the simulations ($\sim 10^{5}$) is
strongly limited compared to experiments ($\sim 10^{15}$), which
benefit of very short microscopic ($\tau_0 \sim 10^{-12} {\rm s}$)
compared to laboratory ($t \sim 10^{0-5} {\rm s}$) time scales.
Experiments \cite{Orbach} and simulations \cite{numerics} have shown
that the spatial growth of the spin-spin correlation length $\xi$ is
very slow ($\xi\sim (t/\tau_0)^{0.15T/T_g}$), reaching hardly 5-10
lattice units in simulations. It is therefore very likely that Edwards
Anderson simulations cannot spatially develop neither the kind of
mechanisms that we have proposed here, nor the hierarchy of embedded
length scales which should correspond to the hierarchical
interpretation of the experiments \cite{hierarki,DWhierarchy}. In that
case, the question of `chaos' in the Edwards Anderson spin glass
\cite{numerics} might remain open for some time.

Finally, let us emphasize that, by discussing temperature dependent
effective interactions as a possible origin of the rejuvenation
effects found in experiments, we want to raise the question of a
possible `classical' origin of apparently `chaos-like' phenomena.
Indeed, an important feature of our present results (as well obtained
in \cite{YOSHINO}) is that the `chaotic' effect is mainly found at
short distances (rejuvenation), while long distance correlations are
preserved (memory). This is very similar to the theoretical case of an
elastic line in presence of pinning disorder \cite{DWhierarchy}, in
which rejuvenation (and memory) effects can also be expected, due to
the selection of {\it smaller and smaller} reconformation length
scales as the temperature is {\it lowered} (see the discussion of
experiments on spin glasses and disordered ferromagnets in
\cite{ferro}). In such a system, when the temperature is decreased,
the small length scales are driven out of equilibrium because of the
{\it classical} thermal variation of the Boltzmann weights of
configurations which were equivalent at a higher temperature,
therefore new equilibration processes must restart.

\section{Conclusion}

In this paper, we have discussed some basic mechanisms which should
be at play in {\it frustrated} (conflicting interactions) and {\it
inhomogeneous} (interactions of various stren\-gths) magnetic systems,
and can thus be at the origin of the so-called `rejuvenation and memory'
phenomena.

Our first point is that, in the presence of frustration, the effective
interaction between two spins (or domains) can take different values with even
different sign at different temperatures (this same result explains
some reentrance phenomena) \cite{REENT,Seiji}.  As a second point, we
have shown that the {\it inhomogeneity} of the interactions can by
itself explain the memory effects, since regions with stronger
interactions and less frustration will naturally remain fro\-zen for
very long times when the temperature is lowered.  We have recalled
simple examples which can be computed exactly.

Combining these two points, we have shown how rejuvenation and memory
can take place in a simple numerical example.  In the case of {\it
real spin glasses}, many complicated bond structures exist, and are
likely to correspond to different equilibrium spin structures for
reasonably separate temperatures.  Hence we expect that in a real spin
glass the above scenario takes place `spontaneously' between different
temperatures.

In our numerical example, we considered that the system of spins {\it
as a whole} was subjected to rejuvenation when the temperature is changed. 
That is, almost all bonds are changed by the temperature change.
However, it is likely that only a part of the system rejuvenates.
Multiple independent rejuvenation and memory stages, 
which take place at different temperatures (like in Fig.2), may then 
correspond to various embedded regions in the sample.

The ordering mechanisms that we have described, in which frustration
and inhomogeneity play a major role, should be important ingredients
for understanding the spin glass phenomena, at least at the mesoscopic
scale, which is the most important for the observable slow dynamics.
The extension of the present {\it real space} approach to a plain
random bond distribution should offer a link with the {\it phase
space} hierarchical picture, and clarify our understanding of the
astonishing properties of spin glasses.

\vspace{1cm}

\noindent {\it Acknowledgments}

We are grateful to H. Yoshino, J.-Ph. Bouchaud, A. Lema\^itre,
V. Dupuis, D. H\'erisson, E. Bertin and J. Hammann for important
discussions on various aspects of this work. We also want to thank the
Monbusho grant (Grant-in-Aid from the Ministry of Education, Science,
Sports and Culture in Japan), which made possible fruitful contacts
and collaborations for several years.

\vspace{1cm}

\section{Appendix: Temperature variations of effective coupling constants}

In this appendix we show that the effective coupling between spins
which interact by frustrated interactions can show a variety of
temperature dependences.  This idea has been discussed in previous
papers \cite{Seiji}, and shown to be at the origin of reentrant phase
transitions \cite{REENT}.  Here we describe some fundamental
mechanisms for various temperature dependences.

Let us consider the effective coupling constant between spins
$\sigma_1$ and $\sigma_2$ related by a linear chain with $n$ bonds like 
in Fig. 9(a).
We assume, for simplicity, that all the bonds are the same and equal to $J$.
We define the effective coupling $K_{\rm eff}$ by the correlation
function of $\sigma_1$ and $\sigma_2$:
\begin{equation}
\langle \sigma_1 \sigma_2 \rangle = \tanh K_{\rm eff}.
\label{Keff0}
\end{equation}
Tracing out the intermediate spins (${\rm s}_1$,$\cdots$ ${\rm s}_{n-1}$),
we obtain
\begin{equation}
K_{\rm eff}(n,T,J)={1\over2}\left( {1+\tanh^n\beta J\over
1-\tanh^n\beta J}\right)
\end{equation}
where $\beta = 1/k_{\rm B} T$. 
The effective interaction $J_{\rm eff}=K_{\rm eff}/\beta$ 
varies with temperature.

Let us now consider two chains of  different lengths $n$ and $m$, like in
Fig. 9(b). The effective
coupling between the spins $\sigma_1$ and $\sigma_2$ is the sum of the 
contributions of the two chains:
\begin{equation}
K_{\rm eff1}(n,m,T,J,J')=K_{\rm eff}(n,T,J)+K_{\rm eff}(m,T,J')\ .
\label{add-couple}
\end{equation}
If the signs of the two contributions are different, these interactions
suffer frustration.
In Fig.10(a), we show the temperature dependence of
$K_{\rm eff1}$ for $(n=10,J)$ and $(m=11,J'=-J)$ as a function of
$T/J$.  We see that the effective coupling stays constant at low
temperatures.  
By the relation (\ref{Keff0}),
the fact that the effective coupling stays constant 
means that the spin correlation function does not develop 
up to unity. 
That is, the spins do not align completely even at low temperatures.  
Here , we used the same amplitudes for $J$ and $J'$ 
in order to cancel the coupling at low temperatures. But of course we can
use different ones.  Indeed, the example given in Section 3
corresponds to $K_{\rm eff1}(1,2,T,J,-2J)$, where the effective 
coupling changes sign.

Because of the additive
nature of the effective couplings Eq.(\ref{add-couple}), we can arrange
various temperature dependences making use of the step-function like
dependence of $K_{\rm eff1}$.  For example, we can make an effective
bond which is
relevant only in a limited temperature range, by combining $K_{\rm
eff1}(n,n+1,T,J_1,-J_1)$ and $K_{\rm eff1}(n,n+1,T,J_2,-J_2)$ as 

\begin{eqnarray}
K_{\rm eff2}(n,T,J1,J2)= K_{\rm eff1}(n,n+1,T,J_1,-J_1) \nonumber \\
+K_{\rm eff1}(n,n+1,T,J_2,-J_2) \ .
\label{KJJ}
\end{eqnarray}
\noindent
Fig. 10(b) shows $K_{\rm eff2}$ in the case $n=10$, $J_1=J$ and 
$J_2=-1.2J$.

An example of a further complicated case is shown in Fig. 10(c), 
the effective coupling of 
which is made of two contributions of the type of (\ref{KJJ}):
\begin{eqnarray}
K_{\rm eff3}(T)= K_{\rm eff2}(n,T,J_1,-1.2J_1) \nonumber \\
+K_{\rm eff2}(n,T,-2J_1,2.4J'_1) \ .
\label{KJJJJ}
\end{eqnarray}
\noindent
The examples in this appendix show that 
a remarkably wide variety of temperature
dependences can be obtained in this way.

\vskip 1.0cm

\begin{figure} \label{A1}
\center\epsfysize=6cm\epsfbox{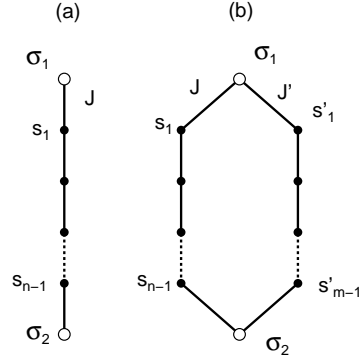}
\caption{Spins $\sigma_1$ and $\sigma_2$ coupled by : 
(a) a chain of $n$ spins 
($s_1,\cdots, s_n$),
(b) two chains of $n$ and $m$ spins 
($s_1,\cdots, s_n$, and $s_1,\cdots, s_m$).}
\end{figure} 

\begin{figure} 
\center\epsfysize=14cm\epsfbox{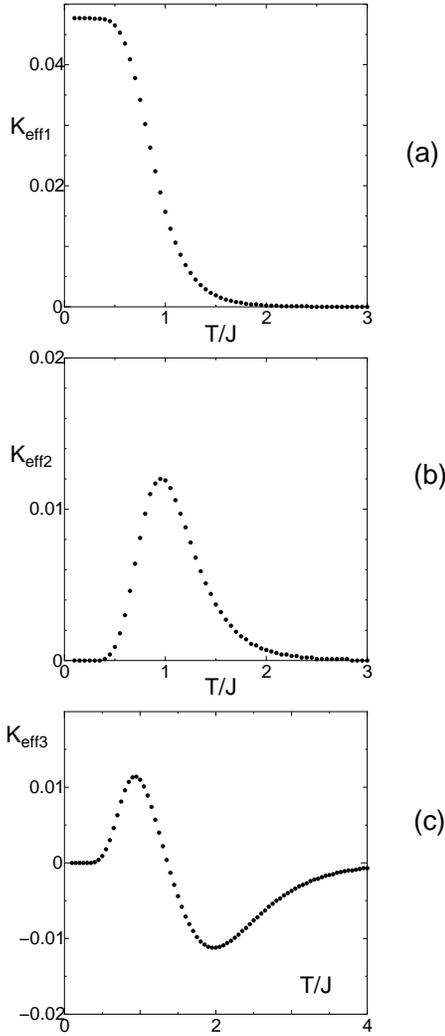}
\caption
{(a) Temperature dependence of $K_{\rm eff1}$ for the case of
$n=10$ and $m=11$ ($J=-J')$.
(b) Temperature dependence of $K_{\rm eff2}$ for the case of
$n=10$ $J_1=J$ and $J_2=-1.2J$.
(c) Temperature dependence of $K_{\rm eff3}$ for the case of
$K_{\rm eff3}(T)= K_{\rm eff2}(n,T,J_1,-1.2J_1)$
$+K_{\rm eff2}(n,T,-2J_1,2.4J'_1)$.}
\end{figure} 

\end{document}